\documentstyle[12pt]{article}
\newcommand{\be}{\begin{equation}}
\newcommand{\ee}{\end{equation}}
\newcommand{\ba}{\begin{eqnarray}}
\newcommand{\ea}{\end{eqnarray}}
\renewcommand{\a}{\alpha}
\renewcommand{\b}{\beta}
\newcommand{\e}{{\bf e}}
\newcommand{\f}{{\bf f}}
\newcommand{\r}{\rho}
\newcommand{\k}{\xi}
\renewcommand{\d}{\partial}

\newcommand{\G}{\Gamma}
\renewcommand{\l}{\lambda}
\newcommand{\s}{\sigma}
\newcommand{\de}{\delta}

\newcommand{\sh}{{\rm sh}}
\newcommand{\ch}{{\rm ch}}
\newcommand{\sign}{{\rm\, sign\,}}
\newcommand{\half}{{1\over 2}}
\newcommand{\g}{{\bf g}}
\newcommand{\x}{{\bf x}}
\newcommand{\ep}{\epsilon}

\renewcommand{\u}{{\bf u}}

\begin{document}
\begin{center}
{\large \bf
INTERTWINING RELATIONS FOR THE MATRIX CALOGERO-LIKE MODELS:
SUPERSYMMETRY AND
SHAPE INVARIANCE.
}\\
\end{center}
\vspace{1cm}
M. V. Ioffe and A. I. Neelov\\ \\
{\small
Department of Theoretical
Physics, University of Sankt-Pe\-ters\-burg, 198504 Sankt-Pe\-ters\-burg,
Rus\-sia.\\ \\
E-mail: m.ioffe@pobox.spbu.ru and neelov@AN6090.spb.edu\\ \\ \\
{\bf Abstract.\ \ }
Intertwining relations for $N$-particle Calogero-like models with internal
degrees of freedom are investigated.
Starting from the well known Dunkl-Polychronakos operators,
we construct new kind of local (without exchange operation) differential
operators. These operators intertwine the matrix Hamiltonians corresponding
to irreducible representations of the permutation group $S_N$. In particular
cases, this method allows to construct a new class of exactly solvable
Dirac-like equations and a new class of matrix models with shape invariance.
The connection with approach of multidimensional supersymmetric quantum
mechanics is established.
}
\section*{\large\bf \quad 1. Introduction}
\hspace*{3ex}

The exactly solvable quantum $N$-body problems
have provided useful tools to investigate both formal algebraic and
analytic properties
with applications to different branches of Physics.
The most intensively studied models are the Calogero model
(many-body extension of the
one-dimensional singular harmomic oscillator)
\cite{calog0}, \cite{calog1} and its various
generalizations, so called Calogero-like models. The latter ones
have either scalar \cite{suther}-\cite{Tur0} or matrix
(with internal degrees of freedom)  \cite{poly2}-\cite{cannata} nature.
The Calogero-like models have been widely developed incorporating
 many-body forces \cite{applications}, different root systems
\cite{Ghosh},\cite{sasaki} and multi-dimensions \cite{multi}.
The supersymmetric extensions of Calogero-like models \cite{Freed}-\cite{mag}
also seem to be
very promising.

 In the papers \cite{dunkl}, \cite{poly1}, \cite{poly2}, \cite{Ghosh},
\cite{Vasiliev0}, \cite{Vasiliev1}, \cite{Vasiliev2} different
types of Dunkl
operators for the investigation of Calogero-like models were
used\footnote{We call the one-particle
operators constructed in \cite{poly1},\cite{poly2} the
Dunkl-Polychronakos operators (DP) to distinguish them from the genuine
Dunkl operators set forth in the
papers \cite{dunkl}, \cite{Vasiliev0}, \cite{Vasiliev1},
\cite{Vasiliev2}, which are slightly different.}.
These operators intertwine Calogero-like Hamiltonians and therefore
allow one to construct the integrals of motion and
the eigenfunctions (if they exist) for these models.

The characteristic trait of DP operators and of the corresponding
integrals of motion derived from them is that they
involve the coordinate exchange operators, and thus
are nonlocal\footnote{By the
words "nonlocal" and "local" we mean "containing exchange operators" and
"containing no exchange operators".}. These techniques are
briefly outlined in Subsection 2.1 of the paper.

On the other hand, the multidimensional supersymmetric quantum
mechanics (SUSY QM) \cite{abi}, applied to the Calogero-like models
\cite{Freed}-\cite{Lap},\cite{sasaki1} provides
one with another set of the
intertwining relations, where the matrix Calogero-like Hamiltonians of a
specific type \cite{mag}, are intertwined by the the supercharge
operators. In
that approach, both the Hamiltonians and the supercharge operators are local.
The intertwining relations are the most important part
of the SUSY QM algebra, which is clear from a number of generalizations
of the standard SUSY QM, e.g. \cite{second},\cite{SIIoffe}.

In the rest of Section 2 these two approaches will be unified:
from the DP operators set forth in \cite{poly1}, \cite{poly2}
we construct new
local operators of the first order in derivatives, that play the role of
intertwining operators between the matrix Calogero-like Hamiltonians. For the
Calogero-Sutherland model \cite{suther} this leads to a new class of
exactly solvable Dirac-like
(matrix and of the first order in derivatives) Hamiltonians. For the Calogero
model with oscillator terms the new intertwining relations give a new
implementation of the shape invariance condition \cite{witten},\cite{Gen},
\cite{eft},\cite{SIIoffe}.

In Section 3 we consider the particular case of three-particle
Calogero-like models,
which is the simplest nontrivial realization of the method
introduced in
Section 2. For the models without oscillator terms (OT) the above mentioned
Dirac-like Hamiltonians coincide with the conventional
Dirac Hamiltonians for a
massless particle in magnetic field. In that case, the $2\times 2$ matrix
Calogero-like Hamiltonians can be interpreted as the
Pauli Hamiltonians for the same system.

In Section 4 the class of the Calogero-like matrix
Hamiltonians described in
\cite{mag} is considered. For such models the intertwining
relations derived in Section 2 are reduced to the well-known SUSY QM
relations \cite{abi}.
However, there is a wide class of models for which the
intertwining relations
introduced in Section 2 are not reduced to any previously known ones.
Clearly, the SUSY QM is valid not only for the Calogero-like models, but for
many other multidimenional and multiparticle ones \cite{abi}. The question as to
how far the generalization of SUSY QM
constructed below can be extended\footnote{ One example of such
extention (for the three-particle case) will be given in
the first footnote on the page 11.} to non
Calogero-like models,  deserves further attention.

A possible way of extension of the formalism presented below is to consider the
generalizations of the Calogero-like models incorporating many-body forces
\cite{applications} and different root systems \cite{sasaki}, for which the
Dunkl operators also exist \cite{Ghosh},\cite{sasaki}, and therefore one can
construct from them local intertwining operators analogous to those of the
present paper.

\section*{\large\bf 2.\quad Intertwining operators of first order
in
derivatives.}
\subsection*{\large\bf 2.1. \quad Dunkl-Polychronakos operators
\cite{poly1},\cite{poly2} for Calogero-like
models.}
\hspace*{1ex}

Let us consider a one-dimensional quantum system of N particles
with coordinates
$x_i$. Let $M_{ij}$ be the operator that exchanges the coordinates
of the $i$-th
and $j$-th particles\footnote{The small latin indices range from 1 to
$N$
everywhere .}. The DP
operators are defined \cite{poly1},\cite{poly2} as:
\ba
\pi_i = -i\d_i + i \sum_{j \neq i} V_{ij} M_{ij}=\pi_i^\dagger
\qquad
V_{ij} \equiv V( x_i - x_j );\label{pi}\\
\d_i\equiv {\d\over \d x_i};\qquad V(x)=V(-x)=V^*(x). \nonumber
\ea
The operators $\pi_i$ are one-particle ones, i.e.
\ba
M_{ij}\pi_i=\pi_jM_{ij};\qquad [M_{ij},\pi_k]=0,\qquad k\ne i,j.
\label{piM}
\ea

Their commutators can be written as:
\ba
[ \pi_i , \pi_j ] = \sum_{k \neq i,j} V_{ijk} [ M_{ijk} - M_{jik}
],\qquad {\rm where}\nonumber\\
V_{ijk} \equiv V_{ij} V_{jk} + V_{jk} V_{ki} + V_{ki} V_{ij},\ \ \ \ \ \ \ \
\ \ \ \ \ \ \ \ \
\ea
and $M_{ijk}$ are the operators of cyclic permutations in three
indices:
\ba
M_{ijk} \equiv M_{ij} M_{jk}= M_{jki} = M_{kij} = M_{jik}^\dagger.
\nonumber
\ea

In the cases both of the Calogero-Sutherland (CS) models
\ba
V(x)&=&l\cot x\qquad {\rm (trigonometric\ or\ TCS\ model ),
}\label{TCS}\\
V(x)&=&l\coth \qquad   {\rm (hyperbolic),}\nonumber
\ea
and of the delta-function model
\ba
V(x)=l\sign x,\label{delta}
\ea
the function $V_{ijk}=l^2$, so that
\ba
[ \pi_i , \pi_j ] = l^2\sum_{k \neq i,j}[ M_{ijk} - M_{jik} ].
\label{pipi}
\ea
The Hamiltonians for these models are \footnote{ The following
procedure is
applicable to arbitrary set of operators $\pi_i$, provided they
satisfy
(\ref{piM}),(\ref{pipi}) and the Hamiltonian is given by the
second equality
of (\ref{H}).}:
\ba
H= -\Delta+\sum_{i\ne j}\biggl[V'_{ij}M_{ij}+V_{ij}^2\biggr]=
\sum_i\pi_i^2+{l^2\over 3}\sum_{i\ne j\ne k\ne i}M_{ijk},\label{H}
\ea
where $\Delta\equiv\sum_i \d_i\d_i$ and $V'_{ij}\equiv V'(x_i-
x_j)$. It is known
\cite{poly1},\cite{poly2} that in this case
\ba
[\pi_i,H]=0. \label{commute}
\ea

In the case of the Calogero model,
\ba
V(x)=l/x; \qquad  V_{ijk}=0;\qquad [\pi_i,\pi_j]=0,\label{Vlx}
\ea
and the equations (\ref{H}),(\ref{commute}) remain valid. What is
more, the DP
operators themselves mutually commute. However, this model doesn't
have a
discrete spectrum.

The Calogero model is usually considered in a harmonic confining
potential (we
abbreviate it as CO: Calogero-Oscillator). For this model, the
following
operators should be introduced \cite{poly1},\cite{poly2}, (see
also
\cite{Ghosh},
\cite{Vasiliev0},\cite{Vasiliev1},\cite{Vasiliev2}):
\ba
a_i^{\pm}=\pi_i\pm \i \omega x_i;\qquad (a_i^+)^\dagger=a_i^-
.\label{ai}
\ea
The Hamiltonian can be written as:
\ba
H_{CO}=\sum_i a_i^+a_i^-+{l\omega }\sum_{i\ne j} M_{ij}
= -\Delta +\omega ^2 \sum_i x_i^2
+\sum_{i\ne j}{l(l-M_{ij})
\over(x_i-x_j)^2}+{N\omega }.\label{Hcal}
\ea
It has been proven \cite{poly2} that the operators $H_{CO}$,
$a^{\pm}_i$
form the oscillator  algebra:
\ba
[H_{CO},a^\pm_j]=\pm 2\omega  a^\pm_j.\label{osc}
\ea

\hspace*{2ex}
\subsection*{\large\bf 2.2. \quad The local form of the
Hamiltonians.}
\hspace*{1ex}

Let us consider an irreducible representation $A$ of the
permutation group $S_N$
realized on real vector functions $f_{\alpha}(x_1,.. .,x_N)$;
$\alpha=1,.
..,\dim A$ by the matrices ${\bf T}^A_{ij}$:
\ba
M_{ij}f_\a= (T_{ij}^A)_{\b\a}f_\beta, \label{zv}
\ea
where $ ( T_{ij}^A) _{\b\a}=(T_{ij}^A) _{\a\b}$ are the (constant)
matrix
elements of the permutation operator $M_{ij}$ in the
representation A. Below we
will assume summation over the repeated indices, unless specified
otherwise. We
will also use the fact \cite{Ham} that ${\bf T}_{ij}^A$ are real
symmetric orthogonal
matrices.

It will be useful to introduce the vector notations:
\ba
{\bf f}= {\bf e}_\a f_\a, \label{vecf}
\ea
where the constant vectors ${\bf e}_\a $ ($\alpha=1,. ..,\dim A$)
form a basis
in
the space of the representation A. Then it is also helpful to
define the
operator ${\bf T}_{ij}^A$ in the vector form:
\ba
{\bf T}_{ij}^A\e_\a =\e_\b(\e_\b)^\dagger {\bf T}_{ij}^A\e_\a \equiv
(T_{ij}^A)_{\a\b}\e_\b.
\label{Tvec}
\ea
Multiplying (\ref{zv}) onto ${\bf e}_\a $ and using (\ref{Tvec}),
we
get:
\ba
M_{ij}\f={\bf T}_{ij}^A\f, \label{zvec}
\ea
where $M_{ij}$ act only on the arguments of $f_\a$ and ${\bf T}_{ij}^A$
only on the
vectors $\e_\a$.

All the Hamiltonians $H$ from the previous Subsection can be
written as
$H=H_{scal}++V'_{ij}M_{ij}$, where $H_{scal}$ are scalar operators
containing no
exchange operator terms:
\ba
H_{scal}=-\Delta+ \sum_{i\ne j} V_{ij}^2 \label{scalfree}
\ea
for the models without  OT (\ref{TCS})-(\ref{delta}), and
\ba
H_{scal}=-\Delta + \omega ^2 \sum_i x_i^2
+\sum_{i\ne j}{l^2
\over(x_i-x_j)^2}+N\omega  \label{scalCO}
\ea
for the CO model (\ref{Hcal}).
The Hamiltonians $H$ act on the functions\footnote{Note that the
functions $\f$
are not necessarily eigenfunctions of $H$. We do not discuss in
the present
paper the symmetry properties of the eigenfunctions of the
Calogero-like
models.}
from the representation $A$ as
\ba
H f_\a=H^A_{\b\a}f_\b;\qquad
H^A_{\b\a} \equiv  H_{scal}\delta_{\a\b}+\sum_{i\ne
j}V'_{ij}(T_{ij}^A)_{\b\a}=
H^A_{\a\b}, \label{HA}
\ea
or, in the vector form,
\ba
H\f={\bf H}^A\f. \nonumber
\ea
Note that if $\f$ satisfies (\ref{zvec}) then $H\f$ satisfies it
too, since Eq. (\ref{zvec}) is equivalent to
the
condition: $ {\bf T}^A_{ij}M_{ij}\f=\f$ (no summation over $i,j$). The
latter
condition
is satisfied for $H\f$, because $[H,{\bf T}^A_{ij}M_{ij}]=0$.

\hspace*{2ex}
\subsection*{\large\bf 2.3.
The intertwining operators in the local form.}
\hspace*{1ex}

The matrix Hamiltonians (\ref{HA}) do not contain exchange
operators $M_{ij}$
explicitly.
Our aim now is to get rid of the $M_{ij}$ in the DP operators
(\ref{pi}),(\ref{ai}) too, and rewrite the Eqs.
(\ref{commute}),(\ref{osc})
in terms of local operators only.

 Let us study the action of the DP operators on the symmetric
functions
satisfying
(\ref{zv}),(\ref{zvec}). The expression $\pi_if_\a$ no longer
satisfies
(\ref{zv}) even when $f_\a$ satisfies it. Instead, $\pi_if_\a$
transforms under
the action of $M_{ij}$ as an object from the direct product of
representations
for $\pi_i$ and $f_\a$. Of course, the DP operators transform
under $S_N$ in
accordance with (\ref{piM}). However, the $\pi_i$ belong to a
reducible
representation of $S_N$ because $\pi_1+...+\pi_N$ realizes the
absolutely
symmetric representation. Therefore, it is helpful to go to the
well-known
Jacobi coordinates \cite{Reed}:
\ba
&y_\k&=\frac{1}{\sqrt{\k(\k+1)}}(x_1+\ldots+x_\k-\k x_{\k+1})
;\qquad 1\le\k\le N-1
\label{yk} \\
&y_N& = \frac{1}{\sqrt{N}}\sum_{i=1}^N x_i, \nonumber
\ea
or\footnote{The indices of the Jacobi variables denoted by Greek
letters range
from 1 to N-1; those denoted by Latin letters range from 1 to N}
$y_k=R_{km}x_m$, where the orthogonal matrix $R$ is determined by
(\ref{yk}).
The derivatives are connected by the same matrix: $\partial /
\partial
y_k=R_{km}\partial / \partial x_m, $ because $R$ is an orthogonal
matrix.
Similarly, we can write the DP operators in the Jacobi variables:
\ba
\r_\k&=&{1\over \sqrt{\k(\k+1)}}( \pi_1+...+\pi_\k-\k\pi_{\k+1})
;\qquad
1\le\k\le N-1
\nonumber\\
\r_N&=&{1\over\sqrt{N}}\sum_{j=1}^{N}\pi_j=-
{i\over\sqrt{N}}(\d_1+..
.+\d_N),\nonumber
\ea
or $ \r_k=R_{k m}\pi_m$.
The operators $\r_\k$ now transform under $S_N$ as the
irreducible representation $\G$ with the Young tableau \footnote {
The standard notation \cite{Ham} for a Young diagram containing
$\lambda_i$ cells in the $i$-th line is
$(\lambda_1,\ldots,\lambda_n)$; if
the diagram contains $m$ identical lines with $\mu$ cells, it is
denoted
by $(\ldots,\mu^m,\ldots)$.  } $(N-1,1)$. Similarly to (\ref{zv})
this fact can
be written as:
\ba
M_{ij}\r_\k=(T_{ij}^\G)_{\l\k}\r_\l. \label{zvro}
\ea
This is a property of the Jacobi variables (see e.g. \cite{mag}).

The object $\r_\k f_\a$ transforms under the action of $S_N$ as
the interior
product $\G\times A$ of the representations $A$ and $\G$, or, in
more detail, in
accordance with the formulae (\ref{zv}),(\ref{zvro}), as
\ba
M_{ij}\r_\k f_\a =(T_{ij}^\G)_{\l\k}(T_{ij}^A)_{\b\a}\r_\l f_\b.
\nonumber
\ea
As outlined in the book \cite{Ham}(chapter 7,\S 13), the interior
product $\G\times
A$ contains only the irreducible representations of $S_N$, whose
Young tableaux
differ from the tableau for $A$ by no more than the position of one
cell (but not
necessarily all of them). For example, for the absolutely
symmetric
representation of $S_N$ with the Young tableau $(N)$, obviously,
\ba
\G\times (N)=\G, \nonumber
\ea
and the result does not contain $(N)$.

Let $B$ be some irreducible representation that appears in
$\G\times A$. Then
we can extract its contribution to $\G\times A$ with the help of
the Clebsch-
Gordan coefficients $(\G\k,A\a|B\s)\equiv(\k\a|\s)$:
\ba
g_\s= (\k\a|\s)\r_\k f_\a= D_{\s\a}f_\a;\qquad D_{\s\a}\equiv
(\k\a|\s)\r_\k.
\label{g}
\ea
The resulting function $g_\s$ satisfies the analog of (\ref{zv})
for the
representation $B$:
\ba
M_{ij}g_\s= (T_{ij}^B)_{\de\s}g_\de. \label{zvg}
\ea
This can be checked directly by substituting (\ref{g}) into
(\ref{zvg}) and
making use of the expression\footnote{The expression
(\ref{Clebsch}) (see
\cite{Ham}, formula (5.114)) is actually a necessary and
sufficient condition for $(C\k,A\a|B\s)$ to be a Clebsch-Gordan
coefficient for
arbitrary representations $A,B$ and $C\in A\times B$.}:
\ba
(T_{ij}^\G)_{\l\k} (T_{ij}^A)_{\b\a}
(\k\a|\s)=(T_{ij}^B)_{\de\s}(\l\b|\de),
\label{Clebsch}
\ea
and the fact that ${\bf T}_{ij}$ are hermitean matrices.

On the functions that satisfy (\ref{zv}) the operator $ D_{\s\a}$
acts as:
\ba
D_{\s\a}f_\a&=& D^A_{\s\a}f_\a;\nonumber\\
 D^A_{\s\a}&=& (\k\b|\s)(\r_\k^A) _{\b\a}=(\k\b|\s)R_{\k k}
(\pi_k^A) _{\b\a};
\label{DA1}\\
 (\pi_k^A)_{\b\a}&=& -i\d_k\de_{\b\a} + i \sum_{m \neq k}
V_{km}(T_{km}^A)_{\b\a}.
\label{DA2}
\ea

For the models without OT $[H,\pi_i]=0$, so, $[H, D_{\s\a}]=0$.
Hence, for any
$f_\a$
\ba
H D_{\s\a}f_\a= D_{\s\a}Hf_\a. \label{HD}
\ea
For all $f_\a$ which satisfy (\ref{zv}), $H^A_{\b\a}f_\b$
satisfies (\ref{zv})
(see the end of the previous Subsection), and
$D_{\s\a}f_\a$ satisfies (\ref{zvg}). Using these symmetry
properties, we can
expand the sides of the equation (\ref{HD}) as
\ba
{\rm(l.h.s.)}= H^B_{\de\s}D_{\de\a}f_\a=
H^B_{\de\s}D^A_{\de\a}f_\a; \nonumber
\\
{\rm(r.h.s.)} = D_{\s\a}
H^A_{\b\a}f_\b=D^A_{\s\a}H^A_{\b\a}f_\b.\nonumber
\ea
Taking into account that ${\bf H}^A,{\bf H}^B$ are symmetric matrices in the
internal
indices, we can see that on the functions satisfying (\ref{zv})
\ba
H^B_{\s\de}D^A_{\de\b}= D^A_{\s\a}H^A_{\a\b}. \label{comm}
\ea

Similarly to the above, for the CO model from the Eqs. (\ref{osc})
it
follows that $HD^\pm_{\s\a}=\ \ \  =D^\pm_{\s\a}(H\pm\omega)$.
Following the same
route that led us from (\ref{HD}) to (\ref{comm}), we can conclude
that on the
functions satisfying (\ref{zv}),
\ba
H^B_{\de\s}D^{A\pm}_{\de\b}= D^{A\pm}_{\s\a}(H^A_{\a\b}\pm 2\omega
\de_{\a\b}),\label{comw}
\ea
where
\ba
D^{A\pm}_{\s\a}&=&(\k\b|\s)R_{\k j} \bigl[(\pi_j^A)_{\b\a}\pm
i\omega
x_j\de_{\a\b} \bigr];\label{DAw}\\
(\pi_j^A)_{\b\a}&=&-i\d_j\de_{\b\a} + il \sum_{m \neq j}
{(T_{jm}^A)_{\b\a}\over
x_j-x_m}.\nonumber
\ea
Note that all terms in the Eqs. (\ref{comm}),(\ref{comw}) are
local: they
contain no exchange operators $M_{ij}$.

\hspace*{2ex}
\subsection*{\large\bf 2.4.
The operatorial nature of the intertwining relations.}
\hspace*{1ex}

The Eqs. (\ref{comm}),(\ref{comw}) are not yet
operatorial
intertwining relations such as, for example, the SUSY QM ones (see
\cite{witten},\cite{Jun},\cite{abi}), because the former are valid
only on the
functions
that satisfy the symmetry condition (\ref{zv}). On the functions
outside that
class the Eqs. (\ref{comm}),(\ref{comw}), generally speaking, may
no longer
be satisfied.

However, we can prove, that they are satisfied on all functions and
thus are
operatorial intertwining relations, by making use of the following

{\bf Theorem 1}:
{\it Let $A$ be some representation of $S_N$. Let $L_{\a\b}$ be
some linear
differential operator of finite order with the coefficients being
rational
matrix functions of the variables $x_i$, or $\sin x_i, \cos x_i$,
or $\sh x_i,
\ch x_i$ (but not of any two of them simultaneously), singular
at\footnote{By
$\x$ without an index we mean the vector $(x_1,...,x_N)$ with $N$ elements.}
$U=\{\x
|
\exists i,j: i\ne j,\ x_i=x_j\}$ at most. The coefficients are
matrices of
dimension $ {\rm dim}A\times {\rm dim}A$. Then, if}
\ba
L_{a\b}f_\b=0\nonumber
\ea
{\it for all $f_\b$ satisfying (\ref{zv}), then $ L\equiv 0$ as an
operator.}

The proof of this Theorem can be found in the Appendix 1. Using
this Theorem 1
for the difference of the left and right parts of the Eqs.
(\ref{comm}),(\ref{comw}) we can conclude that the latter are
satisfied
operatorially\footnote {The
Theorem 1 cannot be applied to the model with $V(x)=l\sign x$,
because the
Eq. (\ref{comm}) may then contain delta-functions-like singularities at
$U$.}.

In particular, it means that that when the initial representation $A$ coincides
with the
resulting representation $B$, the operators ${\bf D}^A$ are integrals of
motion for
the (trigonometric and hyperbolic) matrix CS models. Therefore,
each CS matrix
model corresponding to a representation $A$ such that $A\in
\G\times A$ has a local
integral of motion ${\bf D}^A$ of the first order in derivatives. An
example of model from this
class will be given in the Section 3. Note that e.g. the models
with $A=(N)$
or $A=(1^N)$ lie outside this class.

 With periodic boundary conditions\footnote{
It means that $x_i\in S\equiv [0,2\pi]$, or ${\bf x}\in S^N$, where $S^N$ is a
torus; $\f(x_1,...,x_i+2\pi,...,x_N)=\f(x_1,...,x_i,...,x_N)$. }on a unit
circle $S$, the Hamiltonians ${\bf H}^A$ for the TCS system are exactly
solvable (see e.g. \cite{Pas}) and have discrete spectrum and finite
dimensional degeneracy of levels. The fact that (\ref{comm}), (\ref{comw})
are satisfied operatorially when $B=A$ allows us to find also the spectrum
and the eigenfunctions of the ${\bf D}^A$ , i.e. the normalizable
functions $f_\a(\x),\a=1,...,\dim A$:  \ba D^A_{\a\b}f_\b=\ep
f_\a.\nonumber \ea

The operators ${\bf D}^A$ for the TCS system may be
considered here as Dirac-like Hamiltonians of first order in derivatives:  \ba
 D^A_{\s\a}=(\k\b|\s)R_{\k j}
\biggl[-i\d_j\de_{\b\a} + il \sum_{m \neq j} \cot(x_j-
x_m)(T_{jm}^A)_{\b\a}\biggr]. \nonumber
\ea
>From the commutation relations (\ref{comm}) it follows
that for a given $A$, the operators ${\bf H}^A$ and ${\bf D}^A$ can be
diagonalized
simultaneously\footnote{See \cite{dau3}; compare also with the SUSY QM
intertwining relations which were used to find a part of spectrum of
Hamiltonians in one \cite{sasaki2}, \cite{Aoyama} or two \cite{SIIoffe}
dimensions.}. In more detail, if $\f_b^{(n)}$ are degenerate eigenstates of
${\bf H}^A$ with energy $E_n$:  \ba {\bf H}^A\f_b^{(n)}=E_n\f_b^{ (n)},
\label{HE} \ea and $b=1,..., J$ ($J$ is the degree of degeneracy), then after
acting by ${\bf D}^A$
on both sides of the equality (\ref{HE}) we see that
${\bf D}^A\f_b^{(n)}$ satisfies
(\ref{HE}) too. Hence\footnote{We assume for simplicity that the
action of ${\bf D}^A$
does not destroy the normalizablilty of $\f_b^{(n)}$.}, there
exists a constant
$J\times J$ matrix\footnote{Note
that the matrix elements $ F^{(n)}_{bc}$ are NOT $\dim A\times\dim
A$ matrices,
but just scalar constants. In other words, they do not affect the
vector
structure of $\f_b^{(n)}$.} $F^{(n)}$:
\ba
{\bf D}^A\f_b^{(n)}= F^{(n)}_{bc}\f_c^{(n)} .\nonumber
\ea

Because ${\bf H}^A$ is hermitean, we can choose $ \f_b^{(n)}$ that
constitute a basis
of periodic vector functions with $\dim A$ components on $S^N$.
Because ${\bf D}^A$ is hermitean, one can check that $ F^{(n)}$ is
also hermitean, and it can be diagonalized by a unitary rotation:
\ba
U ^{(n)}_{ab}F^{(n)}_{bc}U^{(n)*}_{dc}=
\ep^{(n)}_a\de_{ac},\nonumber
\ea
and the functions
\ba
\g^{(n)}_a(\x)= U_{ab}^{(n)}\f_b^{(n)}(\x) \nonumber
\ea
are eigenfunctions of the operator ${\bf D}^A$ with energies $
\ep^{(n)}_a$.
Because $U_{ab}^{(n)}$ are unitary matrices, the functions
$\g^{(n)}_a $
constitute a basis of the periodic vector functions on $S^N$.
Hence, they form a full
set of eigenfunctions of ${\bf D}^A$.


For the CO model the case of $B=A$ is interesting too. The Eqs.
(\ref{comw}) will then take the form:
\ba
H^A_{\de\s} D^{A\pm}_{\de\b}= D^{A\pm}_{\s\a}(H^A_{\a\b}\pm
2\omega \de_{\a\b}),
\label{shw}
\ea
where ${\bf D}^{A\pm}$ are still defined by (\ref{DAw}), and
will be
satisfied operatorially. They are the relations of the oscillator-
like algebra
(similar to (\ref{osc})). In the language of SUSY QM it means
that each CO
matrix model for a representation  $A\in \G\times A$
obeys shape
invariance (SI) in $N-1$ dimensions (because the center of mass motion is
decoupled). Let us
stress that this is the first example of SI in several dimensions
realised by
local operators of the first order in derivatives (cf. the
attempts in
\cite{eft}). The nonlocal SI of the
Calogero model (see the Eq. (\ref{osc})) was described in
\cite{poly1},\cite{poly2}, \cite{Ghosh}, \cite{Vasiliev0},
\cite{Vasiliev1},
\cite{Vasiliev2}; a model with a two-dimensional SI of the second
order in
derivatives was proposed in \cite{SIIoffe}.

\section*{\large\bf 3.\quad Examples with $N=3$.}
\hspace*{3ex}

Assume that $N=3$, and both representations $A$ and $B$  are equal
to
$\G=(2,1)$. Then the functions that satisfy (\ref{zvec}) are:
\ba
{\bf f}=
 \pmatrix{ f_{211} \cr f_{121}}.\nonumber
\ea
where the index $\a=(211),(121)$ enumerates the partitions of the
Young tableau
$(2,1)$ (see \cite{Ham} or \cite{Bor}, the end of chapter 1):
\ba
P^+ f_{211}&=& P^- f_{211}=0; \qquad M_{12} f_{211}=
f_{211};\nonumber \\
P^+ f_{121}&=& P^- f_{121}=0; \qquad M_{12} f_{121}=-
f_{121};\nonumber \\
P^\pm&=&1+M_{12}M_{13}+M_{13}M_{12}\pm(M_{12}+M_{23}+M_{13}).\nonumber
\ea
$P^\pm$ are the projectors onto the symmetric/antisymmetric
representations of
$S_3$.
The matrices of the representation $\G$ for $N=3$ have the form:
\ba
{\bf T} ^\G_{12}=\s_3;\qquad {\bf T}^\G _{23}={\sqrt{3}\over 2}\s_1-
\half\s_3;\qquad {\bf T}^\G
_{31}=-{\sqrt{3}\over 2}\s_1-\half\s_3, \label{Tij}
\ea
where $\s_i$ are the Pauli matrices.
The Hamiltonians for the Calogero-like systems without OT for the
representation
$\G$ have the form (\ref{HA}):
\ba
{\bf H}^\G=-\Delta+\sum_{i\ne j}\biggl[V_{ij}^2 +V'_{ij}{\bf T}^\G
_{ij}\biggr]. \label{H3}
\ea

As to the DP operators in the Jacobi variables, one can check that
$\r_1$ has
the same symmetry as $f_{121}$, and $\r_2$ - the same as $f_{211}$.
The Clebsch-
Gordan
coefficients for $\G\times \G\to \G$ are written in the end of
chapter 7 of the
book \cite{Ham}. Plugging them and the ${\bf T}^\G_{ij}$ from
(\ref{Tij}) into the
definition of ${\bf D}^\G$ (expressions (\ref{DA1}),(\ref{DA2})), we can
conclude
after some algebra that for the Calogero-like models without OT
\ba
{\bf D}^\G={1\over\sqrt{2}}\Biggl[-i\s_3 {\d\over\d y_2}+i\s_1{\d\over\d
y_1}-
\sqrt{2}\s_2 (V_{12}+V_{23}+V_{31})\Biggr],\label{D3}
\ea
where $y_\k$ are the Jacobi variables (\ref{yk}). The Eq.
(\ref{comm}) for
${\bf H}^\G$ from (\ref{H3}) and ${\bf D}^\G$ from (\ref{D3}) is satisfied
operatorially,
i.e. $[{\bf H}^\G,{\bf D}^\G]=0$. In fact, a stronger statement for $N=3$ can
be proven by
direct calculation:
\ba
({\bf D}^\G)^2=\half\biggl[ {\bf H}^\G+ {\d^2\over\d y_3^2} \biggr]+C
,\label{square}
\ea
where $C$ is a real constant. What is more\footnote{ One could
even replace the term $V_{12}+V_{23}+V_{31}$ in (\ref{D3}) by an
arbitrary function $v(y_1,y_2)$. Then the square of ${\bf D}^\G$
would remain the sum of the Laplacian and a momentum independent
$2\times 2$ matrix potential. However, we do not know any cases
when this sum is an exactly solvable Hamiltonian, except for those
given in this text.}, it is true
operatorially  even for the case $V(x)=l\sign x$. In that case we
were unable to prove (\ref{comm}) for arbitrary representation $A$, but for
$N=3$ and $A=\G$ it turns out to be true.

Eq. (\ref{square}) signifies that the operator ${\bf D}^\G$ realizes a
sort\footnote{
The operator (\ref{D3}) can be viewed as a Dirac operator for a
massless fermion
in three dimensions $(y_1,y_2,y_3)$ in the magnetic
field that
does not depend
on $y_3$ and is orthogonal to the axis $y_3$. The component of
the fermion's momentum along the axis $y_3$ should be zero. The
Hamiltonian (\ref{H3})
is then the Pauli Hamiltonian for the same system
\cite{Landau},\cite{Jun}.}
of a "square root" of
${\bf H}^\G$. This, in particular, means that the spectrum and
eigenfunctions of
${\bf D}^\G$ itself can be found easier than in the general case
described in Section
2 of this paper, if the spectrum and eigenfunctions of ${\bf H}^\G$ are
known.

For the TCS model the operator ${\bf D}^\G$ (\ref{D3}) has the form:
\ba
{\bf D}^\G={1\over\sqrt{2}}\Biggl[-i\s_3 {\d\over\d y_2}+i\s_1{\d\over\d
y_1}-
l\sqrt{2}\s_2\Biggl(\cot(\sqrt{2}y_1)+\cot\biggl( -{\sqrt{2}\over
2}y_1+\sqrt{3\over 2}y_2 \biggr)+ \nonumber\\ + \cot\biggl(-
{\sqrt{2}\over
2}y_1-\sqrt{3\over 2}y_2\biggr)\Biggr)\Biggr].\nonumber
\ea

The eigenfunctions of this operator are 2-component column
functions
$\g(y_1,y_2)$:
\ba
{\bf D}^\G\g(y_1,y_2)=\ep\g(y_1,y_2). \label{Dep}
\ea
In this case it follows from (\ref{square}) that $\g$ is also an
eigenfunction
of ${\bf H}^\G$ with energy $E=2(\ep^2-C)$.

Let us prove now that all the eigenfunctions of ${\bf D}^\G$ can be
obtained from the
ones of ${\bf H}^\G$. Let $\f(\x)$ be an eigenfunction of ${\bf H}^\G$ with
energy $E$:
${\bf H}^\G\f=E\f$ and zero total momentum. Then the following
alternative should be
considered:

a)  $\f$ itself is already an eigenfunction of ${\bf D}^\G$, i.e. it
satisfies
(\ref{Dep}). Then the corresponding eigenvalue is:
$\ep=\pm\sqrt{E/2+C}$.

b) $ {\bf D}^\G\f\equiv \u $ and $\f(\x)$ are linearly independent.
Then one can check that ${\bf D}^\G\u=(C+E/2)\f$,
and $\u(\x)$ is also an eigenfunction of ${\bf H}^\G$ with energy $E$.
Thus, the
eigenfunctions of ${\bf H}^\G$, that are not the ones of ${\bf D}^\G$, form
pairs in which
${\bf D}^\G$ transforms each member into another. If $\f$ is
normalizable then $\u$
is too, because
\ba
<\u|\u>=<\f|({\bf D}^\G)^2|\f>=(E/2+C)<\f|\f>\ < \ +\infty.\nonumber
\ea
>From each such pair one can construct two eigenfunctions of
${\bf D}^\G$:
\ba
\f\pm (C+E/2)^{-1/2}\u\nonumber
\ea
 with energies\footnote{The situation $C+E/2<0$ is impossible
because otherwise
the Eq. (\ref{Dep}) would remain valid, but with imaginary $\ep$.
The
operator ${\bf D}^\G$ is hermitean, so $C+E/2\ge 0$.}
$\ep=\pm\sqrt{C+E/2}$.

Thus, when taken in the above form, the sets of eigenfunctions of
${\bf H}^\G+{\d^2\over \d y_3^2}$ and ${\bf D}^\G$ concide.

For the CO model the Hamiltonian (\ref{HA}) for the
representation $\G$ will
have the form:
\ba
{\bf H}^\G=  -\Delta +\omega ^2 \sum_i x_i^2
+\sum_{i\ne j}{l(l-{\bf T}^\G_{ij})
\over(x_i-x_j)^2}+3\omega,\nonumber
\ea
where ${\bf T}^\G_{ij}$ are defined in (\ref{Tij}). The intertwining
operators
(\ref{DAw}) can be rewritten as:
\ba
{\bf D}^{\G\pm}={1\over\sqrt{2}}\Biggl[-i\s_3 \biggl({\d\over\d y_2}\mp
\omega
y_2\biggr)+i\s_1\biggl({\d\over\d y_1}\mp \omega y_1\biggr)-
\nonumber\\ -
\sqrt{2}\s_2\Biggl( {1\over y_1}+ {1\over -\half y_1+
{\sqrt{3}\over 2} y_2 }+
{1\over -\half y_1-{\sqrt{3}\over 2} y_2
}\Biggr)\Biggr].\nonumber
\ea
The operatorial relations (\ref{shw}) of the oscillator-like
algebra
\ba
[{\bf H}^\G,{\bf D}^{\G\pm}]= \pm 2\omega  {\bf D}^{\G\pm} \nonumber
\ea
correspond to the SI
of the matrix $2\times 2$ Hamiltonian
${\bf H}^\G$ in two dimensions (because the center of mass motion is
decoupled).

\section*{\large\bf 4.\quad Connection with the ordinary SUSY QM.}
\hspace*{3ex}

In this Section we will restrict ourselves to the class of
representations with
Young tableaux of the form
\ba
A=(N-n,1^n);\qquad n=1,...,N.\label{G}
\ea
 It was proven in \cite{mag} that for this class of
representations we can
choose a basis $\e_\a$ (see (\ref{vecf}))with the help of the
fermionic
creation/annihilation operators $\psi_i,\psi^+_i;\ i=1...N $ :
\ba
\{\psi_i,\psi_j\}&=&0,\qquad
\{\psi_i^+,\psi_j^+\}=0, \qquad
\{\psi_i,\psi_j^+\}=\delta_{ij};\label{ant}\\
\psi_{i}|0>&=&0;\qquad i,j=1...N;\qquad       <0|0>=1. \nonumber
\ea
It is useful to introduce also the
fermionic analogues $ \phi^+_k$ of the Jacobi variables (\ref{yk})
(see
\cite{mag}):
\ba
\phi^+_k= R_{k m}\psi^+_m;\qquad \phi_k=R_{k m}\psi_m, \nonumber
\ea
where $ R_{k m}$ are defined in (\ref{yk}).
The fermionic Jacobi variables obey anticommutation relations
similar to
(\ref{ant}):
\ba
\{\phi_k,\phi_m\}&=&0,\qquad
\{\phi_k^+,\phi_m^+\}=0, \qquad
\{\phi_k,\phi_m^+\}=\delta_{km}.\label{phiphi}\\
\phi_k|0>&=&0;\qquad k,m=1...N. \nonumber
\ea


 Now we can define the basis\footnote{The fermionic
operators $\phi_N, \phi_N^+$ do not enter into $\e_\a$ because
they correspond to the center of mass degree of freedom which is decoupled.}
$\e_\a$:
\ba
\e_\a=\phi^+_{\a_1}... \phi^+_{\a_n}|0>\equiv | \a_1...\a_n>\qquad
\a_i=1,...,N-
1, \label{eferm}
\ea
where $\a\equiv(\a_1...\a_n)$ is a multiindex with values in the
fermionic
number space, and
\ba
(\e_\a)^\dagger\e_\a &=& < \a_n...\a_1 |
\a_1...\a_n>=1;\nonumber\\
(\e_\a)^\dagger &=&(| \a_1...\a_n>)^\dagger =<0|
\phi_{a_n}...\phi_{a_1}\equiv
< \a_n...\a_1|
\ea
 (no sumation over $\a$ is implied).

 Because of (\ref{phiphi}), it is sufficient to include into the
basis only the
vectors $\e_\a $ with, say,
$\a_1<...<\a_n$, and the summation over $\a$ will be done over
such vectors only.

It was also proven in \cite{mag}, that for the basis (\ref{eferm})
the operator
${\bf T}^A_{ij}$ in the vector form (\ref{Tvec}) can be realized as
\ba
(T_{ij}^A)_{\a\b}\e_\b= {\bf T}^A_{ij}\e_\a= {\bf T}^A_{ij} |\a_1...\a_n>
 =K_{ij}|\a_1...\a_n>, \label{TK}
\ea
where
\ba
 \hat K_{ij}\equiv \psi^+_i\psi_j+\psi^+_j\psi_i-\psi^+_i\psi_i-
\psi^+_j\psi_j+1=
1-(\psi_i^+-\psi_j^+)(\psi_i-\psi_j)=\nonumber\\
 =\hat K_{ji}=(\hat K_{ij})^\dagger.
\label{Kij}
\ea
It follows from (\ref{TK}) that all the Hamiltonians ${\bf H}^A$
(\ref{HA}) with $A$
from the class (\ref{G}) take
the same form in the basis (\ref{eferm}):
\ba
H= H_{scal}+\sum_{i\ne j}V'_{ij}K_{ij}. \label{Hgen}
\ea

Let us consider the intertwining relations (\ref{comm}) for the
Calogero-like
models without OT for $A$ from the class (\ref{G}). From
(\ref{Hgen}),(\ref{scalfree}) it follows that the Hamiltonians
${\bf H}^A, {\bf H}^B$ in
(\ref{comm}) will have the form:
\ba
H=-\Delta+\sum_{i\ne j}\biggl[V'_{ij}K_{ij}+V_{ij}^2\biggr]
.\label{HK}
\ea
One may notice that this Hamiltonian is a particular case of  the
Superhamiltonian
given in \cite{mag}, up to the sign of $V_{ij}$ and an additive
scalar
constant.

Because the Young tableaux for $B$ and $A$ belong to the class
(\ref{G}) and can
differ by no more than the position of one cell (see \cite{Ham},
chapter 7,\S
13), $B$ can either coincide with $A$ or have the form $(N-n\mp
1,1^{n\pm 1})$.

Let us consider the case $B= (N-n-1,1^{n+1})$, for which
we can realize (cf. \cite{mag}) the Clebsch-Gordan coefficients in
the operators
${\bf D}^A$ as:
\ba
(\k\a|\s)=< \a_n...\a_1\k | \s_1...\s_{n+1}>=
<\s_{n+1}...\s_{1}|\k\a_1...\a_n
>.\label{n+1}
\ea
One can check that the Clebsch-Gordan coefficients, defined by
(\ref{n+1}),
satisfy
(\ref{Clebsch}), and therefore they correctly connect the
representations $A=(N-
n,1^n),\G=(N-1,1)$ and $B=(N-n-1,1^{n+1})$. These Clebsch-Gordan
coefficients may
differ from the standard ones (see \cite{Ham}) by an inessential
overall factor.

Now we can express the intertwining operators ${\bf D}^A$ (\ref{DA1}) in
terms of the
fermionic operators defined above:
\ba
{\bf D}^A\e_\b=\e_\s(\e_\s)^\dagger {\bf D}^A\e_\b=\e_\s D^A_{\s\b}=
\nonumber\\=|\s_1...\s_{n+1}><\s_{n+1}...\s_{1}
|\k\a_1...\a_n >
R_{\k k}\biggl[-i\d_k\de_{\b\a}  + i \sum_{m \neq k}
V_{km}(T_{km}^A)_{\b\a}\biggr]= \nonumber\\ =
 R_{\k k}\biggl[ -i\d_k|\k\b_1...\b_n >+ i \sum_{m \neq k}
V_{km}(T_{km}^A)_{\b\a}|\k\a_1...\a_n >\biggr]= \nonumber \\  =R_{\k
k}\phi^+_\k\biggl[ -
i\d_k|\b_1...\b_n >
+i \sum_{m \neq k}
V_{km}(T_{km}^A)_{\b\a}|\a_1...\a_n >\biggr]=\nonumber \\= R_{\k k}\phi^+_\k
\biggl[-
i\d_k + i \sum_{m \neq k} V_{km}K_{km}\biggr]|\b_1...\b_n
>.\nonumber
\ea
So, we can conclude that
\ba
{\bf D}^A= R_{\k k}\phi^+_\k \biggl[-i\d_k+i \sum_{m \neq k}
V_{km}K_{km}\biggr].
\label{DKpsi}
\ea
The operator ${\bf D}^A$ turns out not to depend on the specific choice
of $A$ from
the class (\ref{G}), i.e. on the fermionic number. It augments the
fermionic
number by 1; that is related to the fact that ${\bf D}^A$ changes the
position
of one cell in the Young tableau for $A$ (see \cite{mag}).

One can notice that
\ba
R_{\k k}\phi^+_\k = \psi^+_k-{1\over N}\sum_m \psi^+_m,
\label{Rphi}
\ea
and after some algebra one can deduce from (\ref{DKpsi}):
\ba
{\bf D}^A= i\phi_N^+{\d\over\d y_N}+ \psi^+_k\biggl[-i\d_k+i \sum_{m
\neq k}
V_{km}K_{km}\biggr].\label{DK}
\ea
To further simplify the form of the intertwining operator ${\bf D}^A$,
one can use the
following

{\bf Theorem 2:} {\it For all $V_{km}$ with $V_{km}=-V_{mk}$ the
following
equation is satisfied:
\ba
\sum_{m\ne k}\psi^+_lV_{km} K_{km}= \sum_{m\ne
k}\psi^+_lV_{km},\nonumber
\ea
where $K_{km}$ is the fermionic permutation operator (\ref{Kij})}.

The proof of the Theorem 2 can be found in the Appendix 2.

Now one can rewrite the Eq. (\ref{DK}) as:
\ba
{\bf D}^A= -iq^+;\qquad
q^+\equiv -\phi_N^+{\d\over\d y_N}+ \psi^+_k\biggl[\d_k-\sum_{m
\neq k}
V_{km}\biggr].
\label{qp}
\ea
Therefore, the Eq. (\ref{comm}) with the Hamiltonian (\ref{HK})
takes the
form:
\ba
[H,q^+]=0,\label{Hqp}
\ea
and its hermitean conjugation\footnote{The
Eq. (\ref{Hqm}) could also be obtained in another way: we could
consider
the Eq. (\ref{comm}) with $A=(N-n,1^n)$ but with $B= (N-n+1,1^{n-
1})$.
Using formulae similar to (\ref{n+1})-(\ref{qp}) one can check
that then
${\bf D}^A=iq^-=i(q^+)^\dagger$.
} is:
\ba
[H,q^-]=0;\qquad q^-\equiv(q^+)^\dagger. \label{Hqm}
\ea

The operators $q^\pm$ (\ref{qp}) coincide with the su\-per\-charge
ope\-ra\-tors\footnote{The operator $q^+$ from (\ref{qp}) differs from
the standard supercharge operator for the TCS model \cite{Lap} by the term $-
\phi_N{\d\over\d y_N}$ that cancels the dependence of the supercharge on the
center of mass coordinates $y_N$, $\phi_N$.} $q^\pm$ for the
Ca\-lo\-ge\-ro-like mo\-dels gi\-ven in \cite{mag}, ex\-cept for the sign of
$V_{ij}$, which is de\-ter\-mi\-ned by the sign of the constant $l$ in
(\ref{TCS})-(\ref{Vlx}).  Therefore, if we replace $l$ by $-l$ in the SUSY QM
relations of \cite{mag}, they can be rewritten as:  \ba \{q^-,q^+\}=
H+{\d^2\over\d y_N^2}+C;\qquad (q^+)^2=(q^-)^2=0.  \label{SUSY} \ea The term
${\d^2\over\d y_N^2}$ in (\ref{SUSY}) is unimportant because it commutes with
$q^\pm$ and $H$.

The commutation relations (\ref{Hqp}),(\ref{Hqm}) may be
considered as the SUSY
QM commutation relations, corresponding to the algebra
(\ref{SUSY}) with the
supercharges $q^\pm$ and the Superhamiltonian $ H+{\d^2\over\d
y_N^2}+C $.

We can conclude that for the models without OT the intertwining
relations
(\ref{comm}) for the representations $A,B$ from the class
(\ref{G}) turn into
the relations of SUSY QM \cite{abi},\cite{mag}. For other $A$ and
$B$
(\ref{comm}) can be considered as a generalization of the SUSY QM
intertwining
relations\footnote{ The intertwining relations are the most
important part
of the SUSY QM algebra, which is clear from a number of
generalizations of the
standard
SUSY QM: e.g., \cite{second},\cite{SIIoffe}.}.

Now let us turn to the CO model and the intertwining relations
(\ref{comw})
with $A$ from the class (\ref{G}). From
(\ref{Hgen}),(\ref{scalCO}) it follows
that the Hamiltonians ${\bf H}^A, {\bf H}^B$ in (\ref{comm}) are:
\ba
H= -\Delta +\omega ^2 \sum_i x_i^2
+\sum_{i\ne j}{l(l-K_{ij})
\over(x_i-x_j)^2}+{N\omega }.\label{HcalK}
\ea
This Hamiltonian, analogously to the previous case (\ref{HK}), has
the same form
for all representations $A$ from the class (\ref{G}). However, the
Hamiltonian
(\ref{HcalK}) differs slightly from the corresponding Calogero
Hamiltonian given
in \cite{mag}, as will be explained below.

The intertwining operators ${\bf D}^{A\pm}$ can be treated similarly to
the case
without OT, the only difference being that one should write
$\d_i\mp \omega x_i$
instead of $\d_i$ everywhere. In particular, for the case with $B=
(N-n-
1,1^{n+1})$ the definition (\ref{DAw}) leads to the following
analog of the
formula (\ref{DKpsi}) for ${\bf D}^{A\pm}$ (the same for all $A$ from the
class
(\ref{G})):
\ba
{\bf D}^{A\pm}= R_{\k k}\phi^+_\k \biggl[-i\d_k\pm i\omega x_k+i \sum_{m
\neq k} (x_k-
x_m)^{-1}K_{km}\biggr]=\nonumber\\
 =R_{\k k}\phi^+_\k \biggl[-i\d_k+i \sum_{m \neq
k}\biggl(\pm{\omega \over
N}(x_k-x_m)+ (x_k-x_m)^{-1}K_{km}\biggr)\biggr].\label{DpKpsi}
\ea

Making use of the Eq. (\ref{Rphi}) and of the Theorem 2, we can
obtain
that, similarly to (\ref{qp}):
\ba
{\bf D}^{A+}&=& -iq^+;\qquad
q^+\equiv -\phi_N^+ {\d\over\d y_N}+ \psi^+_k\biggl[\d_k-\sum_{m
\neq k}
W_{km}\biggr]; \label{qpp}\\
W_{km}&=& W(x_k-x_m);\qquad W(x)\equiv {\omega \over N}x+{l\over
x};\nonumber \\
{\bf D}^{A-}&=&-i\tilde q^+;\qquad \tilde q^+\equiv -\phi_N^+ {\d\over\d
y_N}+
\psi^+_k\biggl[\d_k-\sum_{m \neq k} \tilde W_{km}\biggr];
\nonumber \\
\tilde W_{km}&=& \tilde W(x_k-x_m);\qquad \tilde W(x)\equiv -
{\omega \over
N}x+{l\over x}.\nonumber
\ea

Taking into account the formulae (\ref{qpp}),(\ref{HcalK}), we can
rewrite the
Eq. (\ref{comw}) for ${\bf D}^{A+}$ and $A$ from the class (\ref{G}) as:
\ba
[H,q^+]=2\omega q^+;\qquad [H, \tilde q^+]=-2\omega \tilde
q^+,\label{Hqw}
\ea
and its hermitean conjugation\footnote{The Eqs. (\ref{Hqmw}) could also be
obtained in
another way:
we could consider the Eq. (\ref{comw}) with $A=(N-n,1^n)$ but with
$B= (N-
n+1,1^{n-1})$. Using formulae similar to (\ref{n+1})-(\ref{qp}),
(\ref{DpKpsi}),
(\ref{qpp}) one can check that then ${\bf D}^{A+}=i\tilde q^-$, where
$q^-$ is
defined in (\ref{qmw}), and ${\bf D}^{A-}=iq^-$.
}:
\ba
[H,q^-]=-2\omega q^-; \qquad [H,\tilde q^-]=2\omega \tilde q^-
;\label{Hqmw}\\
q^-=(q^+)^\dagger= \phi_N {\d\over\d y_N}+ \psi_k\biggl[-\d_k-
\sum_{m \neq k}
W_{km}\biggr];\label{qmw}\\
\tilde q^-=(\tilde q^+)^\dagger= \phi_N {\d\over\d y_N}+
\psi_k\biggl[-\d_k-
\sum_{m \neq k}\tilde W_{km}\biggr]. \nonumber
\ea

The operators $q^\pm$ from (\ref{qpp}),(\ref{qmw}) are similar to
the
supercharge operators\footnote{The operator $q^+$ from (\ref{qpp})
differs from
the standard supercharge operator for the TCS model \cite{Freed}
by the term $-
\phi_N^+{\d\over\d y_N}$ that cancels the dependence of the
supercharge on the
center of mass coordinates $y_N$ and $\phi_N$.} $q^\pm$ given in
\cite{mag} for the Calogero model with OT, up to a redefinition of
constants.
Therefore, we can construct the following SUSY algebra:
\ba
\{q^-,q^+\}=h; \qquad (q^+)^2=(q^-)^2=0, \label{SUSYCO}
\ea
with the Superhamiltonian $h$:
\ba
h=H+2\omega \psi_k^+\psi_k-H_N;\qquad
H_N=- {\d^2\over\d y_N^2}+\omega ^2y_N^2+2\omega
\phi_N^+\phi_N+C, \label{HN}
\ea
where $H$ is defined in (\ref{HcalK}), and $C$ is a scalar
constant. The term
$H_N$ is unimportant because it commutes with
$q^\pm$ and $H$.
The operators $ \tilde q^\pm$ form an algebra similar to
(\ref{SUSYCO}) but
the sign of $\omega $ in the Superhamiltonian (\ref{HN}) should be
different(cf. \cite{Ghosh2}):
\ba
\{\tilde q^-,\tilde q^+\}=\tilde h;\label{SUtilde}\qquad
(\tilde q^+)^2=(\tilde q^-)^2=0;\ \ \ \ \ \ \ \ \ \ \ \ \ \ \nonumber\\
\tilde h=H-2\omega \psi_k^+\psi_k-\tilde H_N; \qquad
\tilde H_N=- {\d^2\over\d y_N^2}+\omega ^2y_N^2-2\omega
\phi_N^+\phi_N+\tilde
C.\nonumber
\ea
>From the SUSY algebrae (\ref{SUSYCO}),(\ref{SUtilde}) one can
deduce the commutation relations that can be shown to be equivalent
to
(\ref{Hqw}),(\ref{Hqmw}):
\ba
[h,q^\pm]=0;\qquad [\tilde h, \tilde q^\pm]=0. \nonumber
\ea

The Eqs. (\ref{comm}),(\ref{comw}) in the case of
$A$,$B$ from
the class (\ref{G}) are reduced to the ordinary multidimensional
SUSY QM
\cite{abi} for the Calogero-like models \cite{mag},\cite{eft}.
However, for $A$
or $B$ outside that class the Eqs. (\ref{comm}),(\ref{comw})
describe a
generalization of the SUSY QM intertwining relations that has not
been known
before. Clearly, the SUSY QM is valid not only for the Calogero-
like models, but
for many others \cite{abi}. The question as to how far the
generalization of
SUSY QM constructed above can be extended to other, non Calogero-
like models,
deserves further attention.

\newpage
\section*{\large\bf Appendix 1.}
\hspace*{3ex}
In this Appendix, we prove the following

{\bf Theorem 1}:
{\it Let $A$ be some representation of $S_N$. Let $L_{\a\b}$ be
some linear
differential operator of finite order with the coefficients being
rational
matrix functions of the variables $x_i$, or $\sin x_i, \cos x_i$,
or $\sh x_i,
\ch x_i$ (but not of any two of them simultaneously), singular
at  $U=\{\x |\exists i,j: i\ne j,\ x_i=x_j\}$ at most. The coefficients are
matrices of
dimension $ {\rm dim}A\times {\rm dim}A$. Then, if}
\ba
L_{a\b}f_\b=0\nonumber
\ea
{\it for all $f_\b$ satisfying (\ref{zv}), then $ L\equiv 0$ as an
operator.}

{\bf Proof:}
Consider the principal Veyl chamber:
$\{\x:x_1<...<x_N\}$. Every function
defined on this chamber can be continued onto the rest of $\Re^{
\dim A}$ by
using (\ref{zv}). The result will obviously satisfy (\ref{zv}), so
it is
annihilated by $L$. Hence, $L$ annihilates all functions on the
principal Veyl
chamber. The same can be stated about every other Veyl chamber:
$\{\x:
x_{i_1}<...<x_{i_N}\}$. Hence, $L$ annihilates all functions on
$\Re^{ \dim
A}\setminus U$. From the fact that the coefficients of
 $L$ are rational functions of $x_i$, or $\sin x_i, \cos x_i$, or
$\sh x_i, \ch
x_i$, it then follows that they are zero identically.

\section*{\large\bf Appendix 2.}
\hspace*{3ex}
In this Appendix, we prove the following

{\bf Theorem 2:} {\it For all $V_{km}$ such that $V_{km}=-V_{mk}$,
\ba
\sum_{m\ne k} \psi^+_k V_{km} K_{km}=\sum_{m\ne
k}\psi^+_kV_{km},\label{theorem}
\ea
where $K_{km}$ is the fermionic permutation operator defined in
(\ref{Kij})}.

Proof: taking into account the definition (\ref{Kij}), we can
check that
\ba
\psi^+_kK_{km}= \psi^+_k+\psi^+_k\psi^+_m\psi_m
+\psi^+_m\psi^+_k\psi_k.
\label{psiK}
\ea
In the Eq. (\ref{psiK}) no summation over either index is implied.
Substituting (\ref{psiK}) into the left side of (\ref{theorem}) we
see that
\ba
\sum_{m\ne k} \psi^+_kV_{km} K_{km}= \sum_{m\ne k} V_{km} \biggl[
\psi^+_k+
\psi^+_k\psi^+_m\psi_m + \psi^+_m\psi^+_k\psi_k
\biggr]=\nonumber\\ = \sum_{m\ne k}\biggl[V_{km}\psi^+_k+ V_{km}
\psi^+_k\psi^+_m\psi_m+ V_{mk}\psi^+_k\psi^+_m\psi_m\biggr]=
\sum_{m\ne
k}\psi^+_kV_{km}.\nonumber
\ea

\section*{\normalsize\bf Acknowledgements}

This work has been made possible in part by the support provided by grant
of Russian Foundation of Basic Researches (N  02-01-00499).  M.I. is indebted
to DAAD (Germany) and the University of Helsinki for their support. He also
warmly
acknowledges  the kind hospitality and useful discussions with Professors
H.-J.Korsch (Kaiserslautern) and M.Chaichian (Helsinki).

\end{document}